\documentclass[aps,prb,superscriptaddress,twocolumn]{revtex4-1}
\usepackage{graphicx}
\usepackage{amssymb}
\usepackage{amsmath}
\usepackage{bm}
\usepackage{wasysym}
\usepackage{color}
\usepackage[T2A]{fontenc}
\usepackage[utf8]{inputenc}
\usepackage[russian,english]{babel}
\usepackage[normalem]{ulem}

\begin{document}
\title{Two-dimensional semimetal in
HgTe quantum well under hydrostatic
pressure}
\author{V.A. Prudkoglyad}
\affiliation{Lebedev  Physical Institute, 119991 Moscow, Russia}
\affiliation{National Research University Higher School of Economics, Moscow,
	101000, Russia}
\author{E.B. Olshanetsky}
\affiliation{Institute of Semiconductor Physics, Siberian Branch of Russian Academy of
	Sciences, Novosibirsk, 630090, Russia}
\author{Z.D. Kvon}
\affiliation{Novosibirsk State University, Novosibirsk, 630090, Russia}
\affiliation{Institute of Semiconductor Physics, Siberian Branch of Russian Academy of
	Sciences, Novosibirsk, 630090, Russia}
\author{V.M. Pudalov}
\affiliation{Lebedev  Physical Institute, 119991 Moscow, Russia}
\affiliation{National Research University Higher School of Economics, Moscow,
101000, Russia}
\author{N.N. Michailov}
\affiliation{Institute of Semiconductor Physics, Siberian Branch of Russian Academy of
	Sciences, Novosibirsk, 630090, Russia}
\author{S.A. Dvoretsky}
\affiliation{Institute of Semiconductor Physics, Siberian Branch of Russian Academy of
	Sciences, Novosibirsk, 630090, Russia}
\date{\today}
\begin{abstract}
 We report results of systematic measurements of charge transport
properties
of the 20.5nm wide
HgTe-based quantum well in perpendicular magnetic field, performed
under hydrostatic
pressures up to 15.1kbar. 
At ambient pressure transport is well described by the two-band semiclassical model.
In contrast, at elevated pressure, we observed non-monotonic pressure dependence of
resistivity at CNP.
For pressures lower than $\approx9$\,kbar, resistivity grows with pressure, 
in accord with expectations from the band structure calculations 
and the model incorporating effects of disorder on transport in 2D semimetals with indirect 
band overlap.
For higher pressures, the  resistivity saturates and
starts decreasing upon further increase of pressure.
 Above  $\approx14$\,kbar  the resistance and hopping transport
character sharply change, which may indicate formation of the excitonic insulator state.
The data also reveals strong influence of
disorder on transport in 2D electron-hole system with a small band overlap.
\end{abstract}
\maketitle

\section{Introduction.}

Two dimensional (2D) semimetals with coexisting electron and hole subsystems,
attract considerable research interest
due to a rich physics which is not fully understood.
For a long time only one example of such system was known, the two-dimensional
electron-hole system
at the interface in GaSb-InAs-GaSb heterostructures
\cite{chang1980electronic,mendez1984quantized,mendez1985quantum,washburn1985new,
munekata1986densities,washburn1986interaction},
where electrons and holes are spatially separated by an energy barrier. The discovery
\cite{Kvon2008,Kvon2011,Olshanetsky2012}
of a new 2D e-h semimetallic system
in HgTe quantum wells (QW) sparked interest to its studies and revived some
long-standing problems in the physics of interacting low-dimensional
systems as well as posed several new questions.

The HgTe QW becomes semimetallic
when the thickness is greater than $\approx 12$\,nm. The band overlap of about
1\,meV \cite{Olshanetsky2012,Minkov2013}
originates from the built-in strain due to the HgTe and CdTe lattice constant mismatch.
The HgTe-based 2D semimetal is noticeably different from
that for GaSb-InAs-GaSb
in several important aspects: (1) it has much higher hole
mobility and concentration, making the role of holes in transport
much more pronounced; (2) electrons and holes are not spatially
separated, therefore, one can expect more vivid manifestations of the electron-hole
interactions; (3) the system is much more tunable by varying carrier
concentrations in the gated structures, by engineering the QW width, and by
tuning the energy spectrum with external pressure.

An old and famous problem in the physics of interacting electronic systems is
the possibility of excitonic insulator (EI) formation. It is well known
\cite{halperin1968excitonic} that semimetallic or semiconducting system should
be unstable against exciton condensate formation, when band overlap or gap,
E$_G$, is lower than the exciton binding energy, E$_B$. Significant complication
comes from the strong influence of disorder on the EI formation.
Since exciton is formed by oppositely charged particles, electrostatic
potential of disorder produces the same effect on EI as magnetic impurity on
superconductor: it lowers transition temperature and, for high enough
concentration of impurities, suppresses pair condensation completely
\cite{zittartz1967theory,zittartz1968ctransport}. As a result, only a few
materials are considered suitable for observation of the EI state.

By now, several systems were suggested to exhibit the excitonic insulator state
\cite{Neuenschwander1990,brandt1972investigation,du2015gate}
In our previous work \cite{olshanetsky2014metal}, an attempt was made to realize
this state in HgTe QW by tuning  interaction with hydrostatic pressure.
The idea behind the experiment was
to diminish the indirect band overlap by applying pressure
to the intrinsically strained heterostructure. Although in Ref.~\onlinecite{olshanetsky2014metal}
we observed an anticipated significant rise of resistivity
under pressure of $\approx 14.4$kbar at low temperature when the system was
adjusted closely to the charge neutrality conditions ($N_e=N_p$), a number of questions remained unaddressed.

The  most important issue concerns evolution of the observed presumably EI
state with pressure and the magneto-transport properties of the system. Recently,
an alternative explanation of our results was suggested  in Ref.~\onlinecite{Knap2014}.
This paper emphasized important role of disorder in transport properties of the
2D semimetals with a small band overlap. It was suggested, that in the presence
of long-range disorder, the 2D semimetal with equal or nearly equal
concentrations of electrons and holes  should likely consist
of spatially separated electron
and hole puddles. Because of the indirect band overlap, the electron-hole
scattering is mostly phonon-assisted and should freeze out at low enough temperatures.
This should lead to a sharp increase of resistivity near the charge
neutrality point when temperature decreases.

In this paper we report results of the extensive charge transport measurements for HgTe-based QW in perpendicular magnetic field, 
performed in the wide range of hydrostatic pressures, up to 15.1kbar. We significantly expanded the explored parameter space, 
as compared to \cite{olshanetsky2014metal}, and due to this, we traced the evolution of the transport with external pressure, 
temperature and magnetic field. Our current results suggest possible excitonic
phase formation in high pressure range (>13\,kbar) and also reveal strong influence of
disorder on transport in 2D electron-hole system with a small band overlap.

\section{Experimental.}
The Cd$_x$Hg$_{1-x}$Te/HgTe/ Cd$_x$Hg$_{1-x}$Te samples with 20.5nm  wide HgTe-quantum well were grown by modified MBE
technology on (100)-GaAs substrate. Details of the sample fabrication and
structure design
may be found elsewhere \cite{dvoretsky2007growing, Olshanetsky2012, Kvon2008,
Olshanetsky2009}. The sample was lithographically defined as Hall bar in order
to measure all components of the resistivity tensor. A TiAu film gate electrode
deposited atop of the SiO$_2$/Si$_3$N$_4$ gate insulator layer, enabled us to
vary the density of holes and electrons in the QW in the range $(1 \div40)\times10^{10}$cm$^{-2}$, at the rate $8\times 10^{10}$cm$^{-2}V^{-1}$.

Resistivity was measured by conventional four probe AC Lock-in technique.
The current modulation frequency and amplitude were chosen in the range 5-20\,Hz and 1-10\,nA, correspondingly.
The low currents ensured the absence of the electron overheating. Hydrostatic
pressure was applied to the samples placed inside a BeCu cell of the
piston-cylinder type with PES-1 as a pressure transmitting medium
\cite{kirichenko2005properties}.
All pressures were generated and clamped at room temperature. Then the pressure
cell was cooled down to liquid helium temperatures. Pressure values cited below were
determined from the superconducting transition of a tin gauge and refer to the
low temperature conditions. The plate-like samples were oriented with 2D layer
normal to the magnetic field direction. All magnetoresistance data presented below for diagonal and off-diagonal components of magnetoresistivity (magnetoconductivity) tensor,
was symmetrized and antisymmetrized correspondingly by taking measurements in both magnetic field directions.
\section{Results and Discussion.}
\subsection{Zero magnetic field resistivity.}
\begin{figure}[t]
\includegraphics[width=8.5cm]{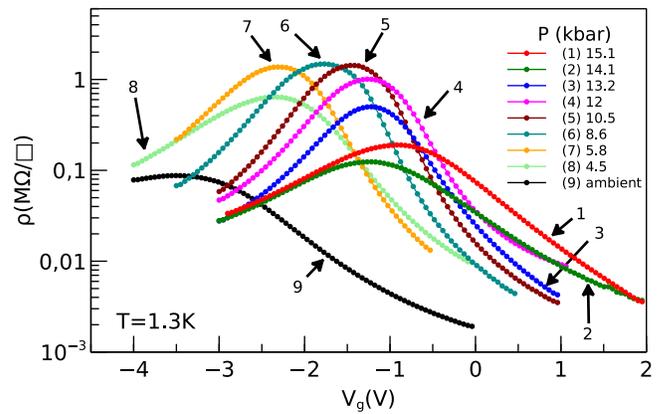}
	\caption{\label{figvgp}(Color online) $\rho$ vs. gate voltage curves  for different
pressure values at $T=1.3$\,K. Ambient pressure data represent sample state after full release of pressure.
}
\end{figure}

Figure~\ref{figvgp} summarizes pressure evolution of the $R(V_g)$ characteristics,
measured at zero magnetic field and at $T=1.3$\,K. Each curve was obtained by
varying gate voltage at fixed temperature and pressure.
For a fixed pressure, the maximal resistivity is achieved when the
concentrations of
electrons and holes are nearly equal; this is called ``charge neutrality point''
(CNP). The maximum is caused by two factors: when the system is close to CNP,
concentrations of charge carriers are low. In addition, electron and hole
scattering  is stronger, because of the increased screening length. This type of
dependence is well known and typical for 2D semimetallic system under
consideration\cite{Olshanetsky2009,Kvon2011,Olshanetsky2012,Gusev2012}.

In principle, one could determine more precisely the actual position of CNP from electron and
hole concentration dependence on gate voltage, but in our case this was not
possible for the reasons given below in the  subsection ``Magnetoresistance measurements''. For low enough temperatures, true CNP and
point of maximal resistivity are very close to each other, their minor difference is insignificant for our purpose,  and in what follows we
will refer to a point of maximal resistivity as CNP with no additional
comments.

As shown in our earlier
work \cite{olshanetsky2014metal} the main difference in the $R(V_g)$
characteristics under ambient and elevated pressure is that the resistivity near CNP sharply increases with pressure. New data reported here, indeed,  confirms  this   previous result. Here we extend our measurements to  pressures up to 15.1\,kbar
in order to trace  systematically changes in the sample behavior with applied pressure. Measurements
were made at the following sequence of applied pressure values:
 0$\rightarrow$
 15.1$\rightarrow$14.1$\rightarrow$12$\rightarrow$13.2$\rightarrow$10.5
$\rightarrow$8.6$\rightarrow$5.8$\rightarrow$4.5$\rightarrow$ 0\,kbar.

When pressure rises up to 8.6\,kbar, sample resistivity also rises. This is the type of behavior one expects from the band structure calculations \cite{krishtopenko2016pressure}, which predict decrease in the band overlap when pressure increases.
Resistivity behavior for higher pressures, however, is quite unexpected: as pressure rises above 8.6\,kbar, the resistivity saturates and starts
decreasing (see Fig.~\ref{figvgp}), indicating significant deviation from the simple band structure model.

\begin{figure}[t]
	\includegraphics[width=8.5cm]{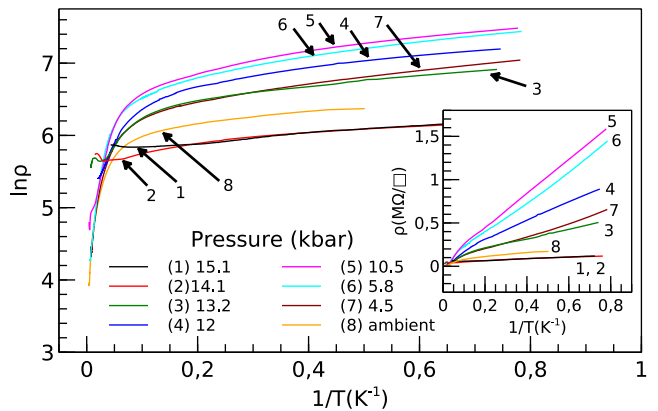}
	\caption{\label{figrtall}(Color online) Temperature dependencies of sample resistivity
at various pressures for gate voltages corresponding to CNP. Ambient pressure data represent sample state after full release of pressure. The inset
shows $R$ vs. $1/T$ dependencies in the low temperature region. Curve numbers on the
inset correspond to the same pressure values as on the main picture.
}
\end{figure}

As was described in introduction, the most interesting physics is expected near
CNP,  therefore we measured temperature dependencies of resistivity at CNP for different
pressure values. Figure~\ref{figrtall} demonstrates $\log \rho$ vs $1/T$
dependencies for the same set of pressures as on Fig.~\ref{figvgp}. Gate voltage
for each curve   was set  at the charge neutrality point  and remained fixed during the temperature run.

The sheet resistance in Fig.\ref{figrtall} typically exceeds
25\,kOhm/$\square$
and one might anticipate a temperature activated or hopping type transport.
This is indeed the case for high temperatures, $T>10$\,K,
however, for low
temperatures the resistivity
versus $1/T$ surprisingly slows down and follows $\rho\propto T^{-1}$ law,
rather than  conventional
hopping-type
$\rho \propto \exp(T_0/T)^p$ dependence \cite{shklovskii2013electronic}.

The slow temperature dependence $\rho(T)$ is  similar to that observed in a more narrow, 8--8.3\,nm, HgTe quantum wells  \cite{Gusev2013a}.
In those structures saturation of the temperature dependence of resistivity at low temperatures was interpreted as a signature of edge transport.
Although we didn't explore other manifestations of edge-channel contribution to conductivity in our quantum wells,
this explanation of the low temperature resistivity ``slowing'' seems plausible.
We note, that
the $1/T$-  type of the $\rho(T)$ dependence  was  observed earlier \cite{Kvon2011} for semimetallic HgTe
QW, grown on (112)-GaAs surface and remained unexplained. If one forces  this dependence to fit  temperature activated transport,
$\rho \propto \exp{\Delta/T}$, then the respective gap  appears to be too small
$\Delta <0.5$K to have a physical sense. For higher temperatures, $T>10$K, (see the main panel
in Fig.\ref{figrtall}) the
$ \ln\rho$ versus $(T_0/T)$
is almost linear and $T_0$  is of the  order of 100\,K.

\subsection{High temperature hopping transport}
\begin{figure}[t]
	\includegraphics[width=8.5cm]{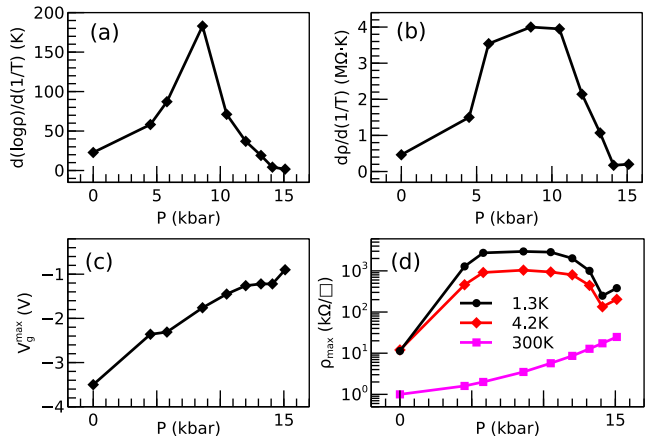}
	\caption{\label{figdiscr}(Color online) (a) Pressure dependence of the slope of $\ln R$
vs. $1/T$ curves, measured at CNP for 15$<$T$<$50\,K. (b) Pressure dependence of low temperature slopes for R vs. 1/T curves. (c) maximum position for R vs. Vg curves at T$=1.3$\,K. (d) Pressure dependence of sample resistivity at CNP for $T=1.3$, 4.2 and 300\,K. Symbols are the data and curves are guide to the eye.
}
\end{figure}

Interestingly, the slopes for both types of dependencies (at low $T$ and at high $T$) in appropriate coordinates
and the value of resistivity at low
temperatures as well, change with pressure in a non-monotonic manner.
These  changes are illustrated by Fig.\ref{figdiscr}, which shows pressure dependencies of the slope,  $d\ln\rho/d(1/T)$, maximum position ($V_{g}^{\rm max}$) on $R(V_g$) characteristics,  and resistivity at
CNP ($\rho_{\rm max}$). Band structure calculations \cite{krishtopenko2016pressure} predict monotonic decrease of the band overlap with pressure. Correspondingly, one should expect a monotonic  increase of resistivity at CNP with pressure. The observed nonmonotonic dependencies fall out of this semiclassical picture and we believe it  results from the interplay of the band overlap and disorder, both parameters vary when pressure changes.

Pressure increase causes
diminishing band overlap and, as a result, effective disorder becomes more important and leads to spatially inhomogeneous state.  The inhomogeneous state may be viewed to consist of spatially separated conduction ``lakes'' and insulating ``barriers'';
the activation energy  $\Delta$ in the temperature activated transport  at high temperature therefore  may be roughly related with a potential barrier height which the carriers need
to overcome for hopping between the neighboring conduction lakes.
In section  \ref{hopping_transport}  further we will discuss the low-temperature ``coherent'' transport data which supports this conjecture of inhomogeneous state consisting of
conduction lakes separated by potential barriers.

With this assumption, we conclude from Fig.~3a that the barrier height grows with pressure up to $\approx 10$\,kbar.
Upon further pressure increase, $\Delta$ falls down to  almost zero and tends to saturate, consistent with the power-low type dependence. The latter, in its turn,  might indicate that the conduction areas start overlapping causing the barrier height to decrease.
Finally, at 15\,kbar the resistance  sharply increases by a  factor of 2-3 (see Fig.~3d). The latter increase might be caused by formation of  EI;
no such anomaly is seen at room temperature (see Fig.~3d, lower curve).

\subsection{Magnetoresistance measurements}
\subsubsection{Overall behavior}
To gather more information on the
origin of qualitative transformations of the charge transport
 with pressure, at each pressure point we performed a series of
magnetoresistance (MR) measurements. The measurements were carried out for gate voltages
ranging from -1.5\,V to +1.5\,V (relative to CNP), in the temperature interval from
1.3 to 30\,K and in magnetic fields up to 4\,T.
The overall picture of magnetotransport looks more clear in terms of conductivity components, $\sigma_{xx}$ and $\sigma_{xy}$, (rather than resistivity) which were obtained from resistivity tensor inversion. This will be further referred to as ``magnetoconductivity''(MC).
We first consider the MC behavior at a particular representative pressure
$P=10.5$\,kbar.

\begin{figure}[h]
	\includegraphics[width=8.5cm]{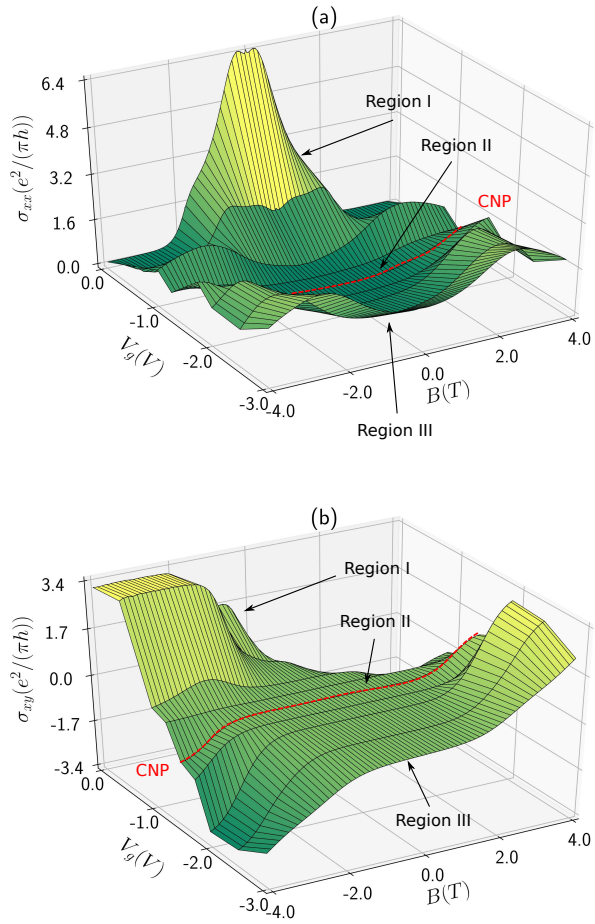}
	\caption{\label{fig3d}(Color online) $\sigma_{xx}$ and $\sigma_{xy}$ dependence on gate voltage, $V_g$, and magnetic field, $B$, for $P=10.5$\,kbar and $T=1.3K$.
}
\end{figure}

The overall picture of МС at this pressure
is shown on Fig.~\ref{fig3d} for the lowest available temperature 1.3\,K.
For the sake of
convenience we divide the examined range of gate voltages into three regions as follows: (I) region of positive gate voltages relative to CNP, where electrons
dominate in the charge transport, (II) region near CNP,  and (III) region of negative voltages, where
contribution to transport from holes is more significant than in region (I).

\begin{figure}[t]
	\includegraphics[width=8.5cm]{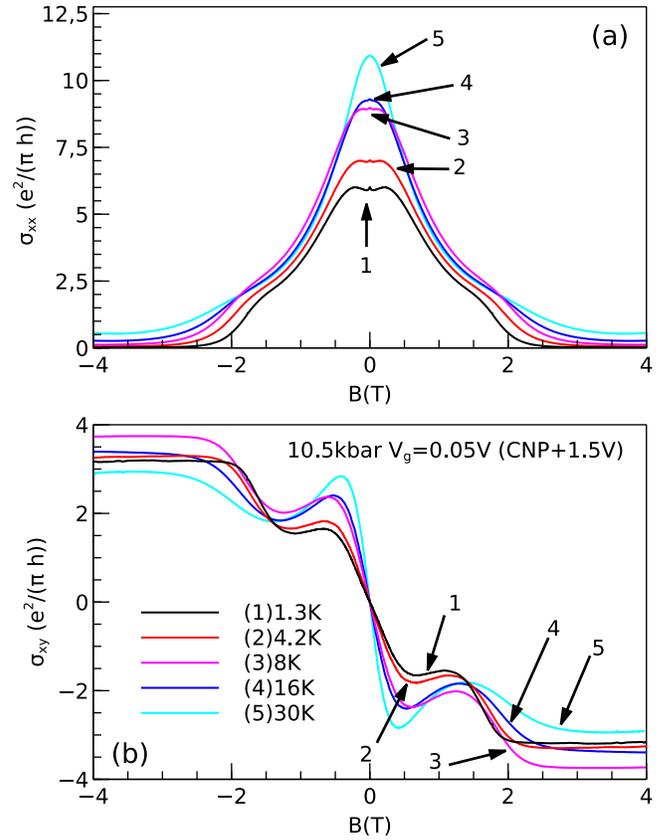}
	\caption{\label{figeside}(Color online) Magnetoconductivity of the sample for
		P=10.5kbar and $V_g=0.05V$ (+1.5V relative to CNP) for a set of temperatures:
		1.3;4.2;8;16;30K. (a) $\sigma_{xx}(B)$, (b) (c) $\sigma_{xy}(B)$}
\end{figure}

Region (I) is characterized by relatively high conductivity in low magnetic field and signatures of the developing quantum hall effect (QHE) in higher field (Fig.~\ref{figeside}).
The diagonal conductivity $\sigma_{xx}(B)$ is negative for all  magnetic fields, whereas  off-diagonal $\sigma_{xy}$ exhibits emerging quantum Hall plateaus corresponding to $\nu=1$ filling factor.
Figure \ref{figeside} shows temperature evolution of the diagonal
(panel (a)) and Hall (panel (b)) magnetoconductivity for the gate voltage $V_g=0.05$\,V (i.e., +1.5\,V relative to CNP).
On Fig.~\ref{figeside} one can also see
signatures of weak localization (WL) and  weak antilocalization (WAL) in low fields $B\ll 0.1$\,T. They quickly vanish as $T$ increases above 1.3\,K.
 (example of weak field MC fit with Hikami-Larkin-Nagaoka formula see on Fig.~13 in Appendix).

\subsubsection{Smooth background semiclassical MC}
Besides  quantum corrections, the MC has a smooth background, which persists throughout the examined temperature range in low magnetic fields (Fig.~5\,a). Although the background MC presumably is of a semiclassical origin, temperature dependence of electron mobility, determined from the best fits of our data with two-band semiclassical model, contradicts the common sense arguments:
the mobility grows with temperature, opposite to expectations based on usual scattering mechanisms.
Would one attempts to explain these features by the CNP shift with temperature,
the electron mobility would have dropped with temperature increasing. However, the  fit with two-band model (for more detail, see Appendix)  does not show such behavior of mobility and
therefore this assumption does not make the two-band model relevant to our data.
Moreover, the $\mu_n(T)$ dependence also  disagrees with the direct experimental data \cite{Olshanetsky2012}.
All these inconsistencies suggest that the two-band model, that
successfully describes
transport at ambient pressure \cite{Olshanetsky2012}, is inadequate at
high pressures.

\begin{figure}[t]
	\includegraphics[width=8.5cm]{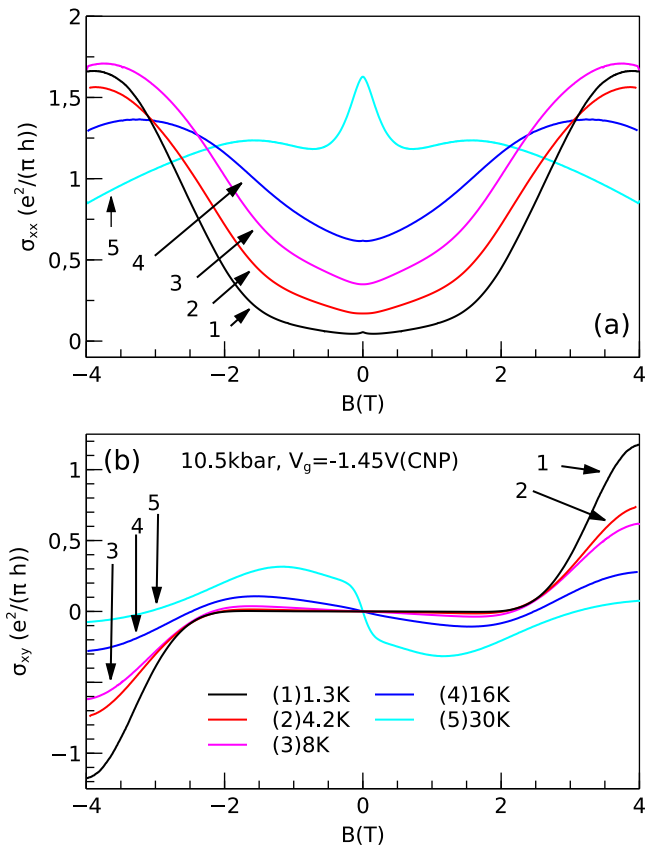}
	\caption{\label{figcnp}(Color online) Magnetoconductivity of the sample for P=10.5kbar
		and $V_g=-1.45V$ (CNP) for a set of temperatures: 1.3;4.2;8;16;30K.
	}
\end{figure}

When we tune gate voltage close to CNP (region (II)), where concentrations of electrons and holes are equal  (see Fig.~4), the overall picture changes drastically, as presented on Fig.~\ref{figcnp}. The diagonal conductivity falls down almost by an order of magnitude at $T=1.3$K in comparison to that at $V_g=0.05V$, thus driving the system to the hopping conduction regime (see further for more detail). This regime corresponds to the wide ``valley'' of low conductivity in Fig.~\ref{fig3d}(a).
The  most striking is that the  off-diagonal component of conductivity
at low temperatures vanishes and remains zero in an extended range of fields   (Fig.~\ref{fig3d}(b)). This effects persists in the interval of pressures from 4.5 to 12kbar and is missing at ambient pressure.

The zero Hall effect was observed earlier for a compensated electron-hole system in InAs/GaSb heterostructure \cite{daly1996zero} in the quantum Hall effect regime.
In that system, the zero Hall state emerges
when the numbers of occupied electron and hole Landau levels are equal. This effect is concomitant of the strong peaks of $\sigma_{xy}$  between the quantized plateaus.
The current situation is very different from the one observed in InAs/GaSb superlattice. In our case the low field magnetotransport
is not determined by the  energy spectrum quantization and certainly is of different origin. We believe that the low diagonal conductivity value, $\sigma_{xx}\ll \sigma_{xy}$, is responsible for the seemingly unusual behavior of $\sigma_{xy}$, which is simply a consequence of the fast growth of $\rho_{xx}$ as temperature decreases.

\subsubsection{Hopping transport features}
\label{hopping_transport}
\begin{figure}[t]
	\includegraphics[width=8.5cm]{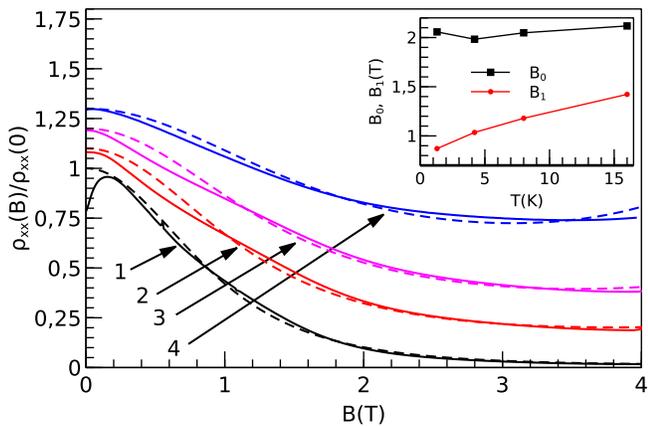}
	\caption{\label{rgfitt}(Color online) Diagonal MR of the sample for P=10.5kbar and
		$V_g=-1.45V$ (CNP) for a set of temperatures: (1)1.3;(2)4.2;(3)8;(4)16K. Solid
		lines show original data,  while dashed ones show results of theoretical
		modeling using formula by Raikh and Glazman\cite{raikh1995negative}. For clarity,
all data are normalized, and
curves (2),(3) and (4) are shifted by 0.1; 0.2 and 0.3, respectively. Corresponding values of $\rho_{xx}(0)$ are as follows: 1880k$\Omega/\Box$ for 1.3K, 486k$\Omega/\Box$ for 4.2K, 234k$\Omega/\Box$ for 8K and 132k$\Omega/\Box$ for 16K. 		Inset shows  temperature dependence of the model parameters.
}
	\end{figure}

The second major feature of $\sigma_{xx}(T,B)$ is that the low field magnetoconductivity in region II is positive (PMC), as opposed  to that for region I;  PMC persists  up to, at least 16\,K (Fig.6a). To understand this behavior we temporarily switch back from conductivity components to resistivity. The magnetoresistance data is shown on Fig.~\ref{rgfitt} by the solid lines. As one can see, resistivity  here much exceeds 26k$\Omega/\Box$ which is the upper limit for diffusive transport regime.
Application of perpendicular magnetic field causes resistivity decrease by a factor of 50 at $T=1.3K$ and $B=4T$.
As temperature increases, the  negative magnetoresistance $\delta\rho(H)/\rho(0)$ gets smaller, but still  is of the order of unity.
For instance,  $\delta\rho(H)/\rho(0)$ changes by more than $30\%$ at $T=16$K (Fig.~7).

The strong negative magnetoresistance is known to be intrinsic to the hopping conduction regime.
A relatively simple model for negative MR of 2D disordered systems
was proposed
by Raikh and Glazman\cite{raikh1995negative}. This model assumes carrier hopping
between regions with delocalized states; it  predicts quadratic fall of
resistivity in low magnetic fields and exponential growth for higher fields.
Such picture is feasible for 2D semimetal with small band overlap in the
presence of impurity potential, as was pointed out in \cite{Knap2014}, it was also observed for 2D electron system in the hopping regime \cite{voiskovskii1995negative} and seems applicable to our case.
The analytical
formula, describing resistivity behavior, with a number of simplifying
assumptions is
\begin{equation}
\frac{R(B)}{R(0)}=\exp\left(\frac{B^2}{B_0^2}\right)\frac{1}{\cosh^2(B/B_1)},
\end{equation}
where B$_0$ is connected to the parameters of potential barrier separating
``lakes'' of delocalized states, and B$_1$ is related to geometry of the lakes being roughly
 inversely proportional to the average lake area.
Fitting our data with this model and using two adjustable parameters is shown on Fig.~\ref{rgfitt} by dashed lines.
One can see a satisfactory qualitative agreement of the model with experimental data.

Although in the original paper \cite{raikh1995negative} the authors do not
analyze temperature dependence of negative magnetoresistance,
a brief comment is given there, that when the phase coherence length becomes comparable or shorter
than the average lake size, one should expect the field dependence of resistivity to weaken.
This is also in line with our observations. Temperature dependencies for the
model parameters $B_0$ and $B_1$, obtained from fitting, are shown on the insert to
Fig.\ref{rgfitt}.
While $B_0$ is almost temperature independent, growth of $B_1$
indicates squeezing  of the conducting lakes with temperature. The latter seems
consistent with a simple picture of the lake, as a shallow potential well, where
the number of bound carriers decreases as temperature becomes comparable with
the confining potential.

If one considers the  negative (-1.5\,V, relative to CNP) voltage biases (region(III)), the conductivity  increases,
but still corresponds to the hopping type transport. As well as in region II, the  magnetoconductivity $\sigma_{xx}(B)$
is well fitted with Glazman and Raikh model \cite{raikh1995negative}.

\begin{figure}[t]
	\includegraphics[width=8.5cm]{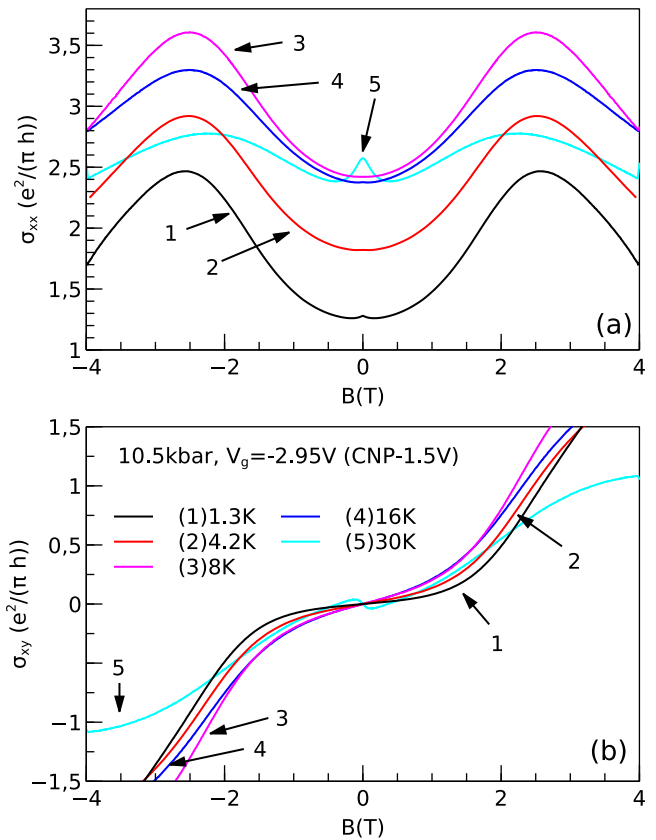}
\caption{\label{fighside}(Color online) Magnetoconductivity of the sample for
$P=10.5$kbar and $V_g=-2.95V$ (-1.5V relative to CNP) for a set of temperatures:
1.3; 4.2; 8; 16; 30\,K. Inset in panel (a) shows results of smooth background subtraction from data. The curve for T=30\,K is divided by 4 to match the scale.}
\end{figure}

\subsubsection{Weak antilocalization and puzzling MC in weak fields}
In weak fields $B<0.1$T and at the lowest temperatures, the  narrow
peak is observed both in the vicinity of CNP  and far away of it, i.e. in regions I, II and III (see Figs.~5a, 6a, and 8a).
Its width and magnitude correspond to the anticipated WAL; this effect was explored earlier \cite{Minkov2012} and is not discussed  here.
The WAL peak gradually disappears as temperature increases to 8K, which is also typical for the WAL effect.

Upon  further increase in temperature
above $\approx 16$\,K, another, {\em wider peak} of negative magnetoconductivity emerges in low fields (see curve 5 in Figs.~5a, 6a, and 8a).
The peak amplitude is much larger than that for the WAL and because of the high temperature it must be of a classical origin. This
 peak is accompanied with a sharp zig-zag change in the  $\sigma_{xy}(B)$ dependence, as can be seen on Figs.~5b, 6b, and 8b.
The slope of the Hall conductivity at $B=0$ changes sign from positive to negative, indicating rise of electronic contribution to transport.
 Surprisingly, the $\sigma_{xx}$ peak amplitude {\em grows} as temperature increases from 16 to 30K.

This unusual behavior is qualitatively consistent
with the predictions given in Ref.~\onlinecite{Knap2014} for temperature evolution of Hall conductivity for negative gate voltages.
According to the theory, as temperature increases, the phonon-assisted electron-hole scattering grows and the electronic contribution to the transport
becomes dominating because the electron mobility much exceeds that of holes.
Physically, the conduction increases at high temperatures due to the increasing probability of carriers scattering between the electron and hole lakes,
and transport obtains a conventional character typical for the  two bands semiclassical transport regime.
As magnetic field increases, the semiclassical two-band conduction produces negative magnetoconductivity; the interplay of these two factors shape the MC peak (curves 5  in Figs.~5a, 6a, and 8a).
Upon further increase of field (e.g., above 0.5T on Fig.~6a)  one can see a  positive MR  which may be interpreted as residues of the hopping-type NMR of the Raikh-Glazman type.
>From the above considereation we conclude
that at  high temperatures two mechanisms, at least, contrbute to transport: (i) lake-to-lake hopping (with Raikh type negative MR), and (ii) electron-hole scattering leading to delocalization and diffusive transport.

To summarize this section, at a fixed pressure
for high carrier concentrations (away of CNP), we observed developing of a conventional   QHE regime  with increasing magnetic field.
For low carrier concentration, we see switching from semiclassical to hopping regime as temperature decreases from 30K to 1.3K.
This behavior in some respects agrees with theoretical predictions, given in Ref.~\onlinecite{Knap2014}, but some observed features don't find explanation within the semiclassical framework.

\subsubsection{Effect of pressure on magnetotransport.}
\begin{figure}[t]
\includegraphics[width=8.5cm]{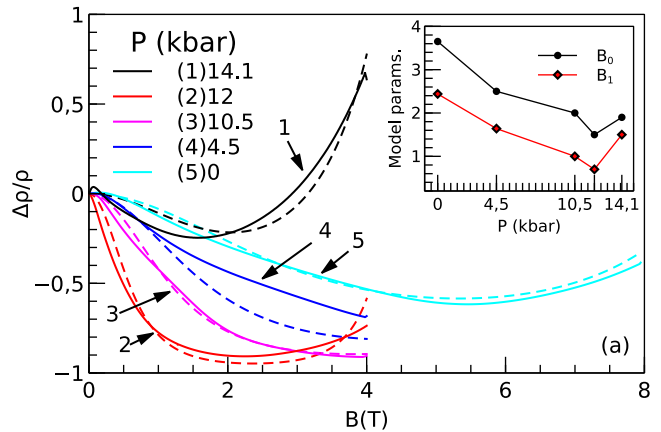}
\caption{\label{figrgp}(Color online) Results of $R(B)$ data fitting with
Raikh-Glazman formula\cite{raikh1995negative} for different pressure values. $V_g$ for all curves is near CNP, T=1.3\,K. The data are shown with solid lines and results of theoretical modeling with dashed ones.}
\end{figure}

Figure~\ref{figrgp} shows how magnetotransport changes with pressure at CNP.
Magnetoresistivity near CNP can be described by the model of Raikh and Glazman\cite{raikh1995negative} for all pressures.  The corresponding model curves and dependencies of model parameters on pressure are shown on Fig.~\ref{figrgp}(a). As one can see, $B_0$ and $B_1$ decrease  as
pressure increases  from zero to $\approx 12$\,kbar. in other words, the electronic and hall lakes
induced by the seeding disorder \cite{Knap2014} grow in size with pressure which may be interpreted as an effective delocalization.
This behavior is consistent with the picture of the band overlap decrease with  increase of pressure.
The behavior of $B_0$ and $B_1$ for higher pressures (P$>12$\,kbar)
sharply changes: both $B_0$ and $B_1$ start increasing thus indicating tendency to the disappearence of lakes and the barriers between them.
Though the  changes of $B_0$ and $B_1$ are not big, they produce an exponentially large impact on the resistivity.
This agrees with sharp changes of the resistivity above 14\,kbar (Fig.~3d). The sharp switching of the hopping transort characterstics above 14kbar may indicate formation
of the EI state.

For pressures lower than 4.5\,kbar,  the low conductivity ``valley'' near CNP is almost absent \cite{supp}.
Magnetoconductivity here is consistent with the hopping transport model  by Raikh and Glazman\cite{raikh1995negative}.

Based on the observed evolution of magnetotransport we can conclude that (1) transport in a system near CNP corresponds to a hopping conductivity regime for pressures up to 14.1 kbar; (2) strong localization of carriers in this state seems to be driven by disorder, although the effect of electron-hole correlations can not be excluded; (3) transport properties near CNP strongly depend on the band overlap, tuned by application of pressure.

\subsubsection{Irreversible effect of high pressure on the QW boundaries }
\begin{figure}[t]
	\includegraphics[width=8.5cm]{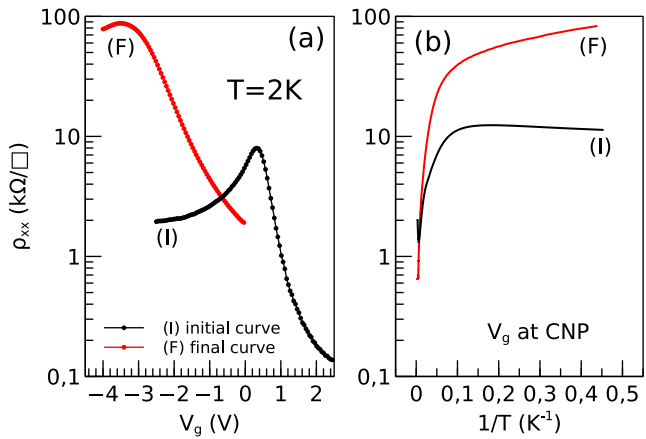}
	\caption{\label{change}(Color online) Comparison of (a)$\rho(V_g)$ and (b)$\rho(T)$ dependencies before and after application of pressure to the sample.
}
\end{figure}

After completion of the course of measurement on the way 15\,kBar $\rightarrow 0$,   when pressure was finally released, we observed an irreversible change of the sample resistivity in comparison to the initial value. This is shown on Fig.~\ref{change}, where initial and final $\rho(V_g)$ characteristics at ambient pressure and $T=2$\,K are plotted.

A detailed analysis of the data showed that the irreversible changes occurred right after application of the 15\,kbar pressure, while further, as pressure was lowered, the sample properties changed quite reversibly.
	
To clarify  the origin of the irreversible change in sample properties after
pressure application cycle we made X-ray analysis of two samples: one exposed to
high pressure and the pristine one. Rocking curve measurements revealed
significant increase of dislocation density in QW after application of 15.1 kbar pressure. Therefore, the factor of ten increase in  the sample resistivity after pressure release can be attributed to the generated dislocations. A concomitant effect of dislocation formation is the diffusion of Hg ions from the quantum well and barrier layers.
The Hg ions are known to easily leave charged vacancies in a lattice.
The structural changes impede disentangling the effects of disorder and
lattice deformation on transport properties of the system. Nevertheless we made sure that behavior of the sample resistivity under pressure was quite reproducible,
i.e. new defects were generated mainly at the initial loading of the cell. So we can attribute the reported changes of the sample resistivity with pressure to
 band structure changes rather than to changes in disorder.
Strictly speaking, the  disorder when compared with the band overlap, inevitably changes also with pressure, thus invoking both parameters to the problem.

\section{Conclusion.}
In summary, our experimental studies showed that application of pressure to a wide
HgTe quantum well drastically alters its transport properties in comparison to ambient conditions.
At ambient pressure transport is well described by the two-band semiclassical model.
In contrast, at elevated pressure, we observed non-monotonic pressure dependence of resistivity at CNP. For pressures lower than $\approx9$\,kbar, resistivity grows with pressure, in accord with expectations from the band structure calculations\cite{krishtopenko2016pressure} and the model incorporating effects of disorder on transport in 2D semimetals with indirect band overlap\cite{Knap2014}.

For higher pressures, the  resistivity saturates and  starts decreasing upon further increase of pressure. Above pressure of $\approx14$\,kbar  the resistance and hopping transport character sharply change, which may indicate formation of the EI state.
 Resistivity at CNP in the high temperature range follows the activated type temperature dependence,
but its temperature changes weaken  at low temperatures, where $\rho(T)$ follows $\rho\sim1/T$ law. This can be compared to the temperature dependence of resistivity for 2D disordered semimetal, calculated in Ref.~\onlinecite{Knap2014}. 
By redrawing it in  logR-1/T and R-1/T coordinates, one can see that at low temperatures it is qualitatively very similar to our data: it saturates at low temperatures and demonstrates nearly linear shape in R-1/T coordinates. Nevertheless, calculated temperature dependence doesn't show exponential growth at higher temperatures and do show saturation of R(1/T) dependence at low enough temperatures.
The activation energy and the resistivity value itself change nonmonotonically with pressure, showing a maximum at $\approx 9$\,kbar; we interpret this in terms of the pressure effect on the conductive lake size and insulating barriers between them.
The resistivity at CNP and at negative gate voltages decrease dramatically with  field at low B, followed by significant growth in higher magnetic field. This behavior is described reasonably well within the framework of Raikh and Glazman model\cite{raikh1995negative} of hopping conduction in the whole range of examined pressures. For temperatures above 16K we observe emergent conductivity peak and sharp sign change of the Hall conductivity in low magnetic fields,  which may be explained as a result of the interplay of two different transport mechanisms, hopping and semiclassical two band diffusion.

As a side effect, we found that  pressure also  introduces defects in the crystal structure and revealed strong dependence of sample transport properties on disorder. Although we did not observe clear evidence of excitonic phase formation near CNP, a dielectric state, present at CNP is of quite unusual type and  some of its unusual transport properties may be find explanation by taking the  electron-hole interaction into account. We believe, further experimental research and development of appropriate theoretical models as well would be essential to understand the nature of possible electronic phases in disordered and interacting 2D semimetals.

\section{Acknowledgements}
We thank S.S. Krishtopenko and V.P. Martovitsky for illuminating discussions. This work has been supported in part by RFBR Grant No. 16-32-00910. VMP acknowledges support by RSF 16-42-01100.

\begin{thebibliography}{29}%
	\makeatletter
	\providecommand \@ifxundefined [1]{%
		\@ifx{#1\undefined}
	}%
	\providecommand \@ifnum [1]{%
		\ifnum #1\expandafter \@firstoftwo
		\else \expandafter \@secondoftwo
		\fi
	}%
	\providecommand \@ifx [1]{%
		\ifx #1\expandafter \@firstoftwo
		\else \expandafter \@secondoftwo
		\fi
	}%
	\providecommand \natexlab [1]{#1}%
	\providecommand \enquote  [1]{``#1''}%
	\providecommand \bibnamefont  [1]{#1}%
	\providecommand \bibfnamefont [1]{#1}%
	\providecommand \citenamefont [1]{#1}%
	\providecommand \href@noop [0]{\@secondoftwo}%
	\providecommand \href [0]{\begingroup \@sanitize@url \@href}%
	\providecommand \@href[1]{\@@startlink{#1}\@@href}%
	\providecommand \@@href[1]{\endgroup#1\@@endlink}%
	\providecommand \@sanitize@url [0]{\catcode `\\12\catcode `\$12\catcode
		`\&12\catcode `\#12\catcode `\^12\catcode `\_12\catcode `\%12\relax}%
	\providecommand \@@startlink[1]{}%
	\providecommand \@@endlink[0]{}%
	\providecommand \url  [0]{\begingroup\@sanitize@url \@url }%
	\providecommand \@url [1]{\endgroup\@href {#1}{\urlprefix }}%
	\providecommand \urlprefix  [0]{URL }%
	\providecommand \Eprint [0]{\href }%
	\providecommand \doibase [0]{http://dx.doi.org/}%
	\providecommand \selectlanguage [0]{\@gobble}%
	\providecommand \bibinfo  [0]{\@secondoftwo}%
	\providecommand \bibfield  [0]{\@secondoftwo}%
	\providecommand \translation [1]{[#1]}%
	\providecommand \BibitemOpen [0]{}%
	\providecommand \bibitemStop [0]{}%
	\providecommand \bibitemNoStop [0]{.\EOS\space}%
	\providecommand \EOS [0]{\spacefactor3000\relax}%
	\providecommand \BibitemShut  [1]{\csname bibitem#1\endcsname}%
	\let\auto@bib@innerbib\@empty
	\bibitem [{\citenamefont {Chang}\ and\ \citenamefont
		{Esaki}(1980)}]{chang1980electronic}%
	\BibitemOpen
	\bibfield  {author} {\bibinfo {author} {\bibfnamefont {L.}~\bibnamefont
			{Chang}}\ and\ \bibinfo {author} {\bibfnamefont {L.}~\bibnamefont {Esaki}},\
	}\href@noop {} {\bibfield  {journal} {\bibinfo  {journal} {Surface Science}\
		}\textbf {\bibinfo {volume} {98}},\ \bibinfo {pages} {70} (\bibinfo {year}
		{1980})}\BibitemShut {NoStop}%
	\bibitem [{\citenamefont {Mendez}\ \emph {et~al.}(1984)\citenamefont {Mendez},
		\citenamefont {Chang}, \citenamefont {Chang}, \citenamefont {Alexander},\
		and\ \citenamefont {Esaki}}]{mendez1984quantized}%
	\BibitemOpen
	\bibfield  {author} {\bibinfo {author} {\bibfnamefont {E.}~\bibnamefont
			{Mendez}}, \bibinfo {author} {\bibfnamefont {L.}~\bibnamefont {Chang}},
		\bibinfo {author} {\bibfnamefont {C.-A.}\ \bibnamefont {Chang}}, \bibinfo
		{author} {\bibfnamefont {L.}~\bibnamefont {Alexander}}, \ and\ \bibinfo
		{author} {\bibfnamefont {L.}~\bibnamefont {Esaki}},\ }\href@noop {}
	{\bibfield  {journal} {\bibinfo  {journal} {Surface Science}\ }\textbf
		{\bibinfo {volume} {142}},\ \bibinfo {pages} {215} (\bibinfo {year}
		{1984})}\BibitemShut {NoStop}%
	\bibitem [{\citenamefont {Mendez}\ \emph {et~al.}(1985)\citenamefont {Mendez},
		\citenamefont {Esaki},\ and\ \citenamefont {Chang}}]{mendez1985quantum}%
	\BibitemOpen
	\bibfield  {author} {\bibinfo {author} {\bibfnamefont {E.}~\bibnamefont
			{Mendez}}, \bibinfo {author} {\bibfnamefont {L.}~\bibnamefont {Esaki}}, \
		and\ \bibinfo {author} {\bibfnamefont {L.}~\bibnamefont {Chang}},\
	}\href@noop {} {\bibfield  {journal} {\bibinfo  {journal} {Physical review
				letters}\ }\textbf {\bibinfo {volume} {55}},\ \bibinfo {pages} {2216}
		(\bibinfo {year} {1985})}\BibitemShut {NoStop}%
	\bibitem [{\citenamefont {Washburn}\ \emph {et~al.}(1985)\citenamefont
		{Washburn}, \citenamefont {Webb}, \citenamefont {Mendez}, \citenamefont
		{Chang},\ and\ \citenamefont {Esaki}}]{washburn1985new}%
	\BibitemOpen
	\bibfield  {author} {\bibinfo {author} {\bibfnamefont {S.}~\bibnamefont
			{Washburn}}, \bibinfo {author} {\bibfnamefont {R.~A.}\ \bibnamefont {Webb}},
		\bibinfo {author} {\bibfnamefont {E.}~\bibnamefont {Mendez}}, \bibinfo
		{author} {\bibfnamefont {L.}~\bibnamefont {Chang}}, \ and\ \bibinfo {author}
		{\bibfnamefont {L.}~\bibnamefont {Esaki}},\ }\href@noop {} {\bibfield
		{journal} {\bibinfo  {journal} {Physical Review B}\ }\textbf {\bibinfo
			{volume} {31}},\ \bibinfo {pages} {1198} (\bibinfo {year}
		{1985})}\BibitemShut {NoStop}%
	\bibitem [{\citenamefont {Munekata}\ \emph {et~al.}(1986)\citenamefont
		{Munekata}, \citenamefont {Mendez}, \citenamefont {Iye},\ and\ \citenamefont
		{Esaki}}]{munekata1986densities}%
	\BibitemOpen
	\bibfield  {author} {\bibinfo {author} {\bibfnamefont {H.}~\bibnamefont
			{Munekata}}, \bibinfo {author} {\bibfnamefont {E.}~\bibnamefont {Mendez}},
		\bibinfo {author} {\bibfnamefont {Y.}~\bibnamefont {Iye}}, \ and\ \bibinfo
		{author} {\bibfnamefont {L.}~\bibnamefont {Esaki}},\ }\href@noop {}
	{\bibfield  {journal} {\bibinfo  {journal} {Surface Science}\ }\textbf
		{\bibinfo {volume} {174}},\ \bibinfo {pages} {449} (\bibinfo {year}
		{1986})}\BibitemShut {NoStop}%
	\bibitem [{\citenamefont {Washburn}\ \emph {et~al.}(1986)\citenamefont
		{Washburn}, \citenamefont {Webb}, \citenamefont {Mendez}, \citenamefont
		{Chang},\ and\ \citenamefont {Esaki}}]{washburn1986interaction}%
	\BibitemOpen
	\bibfield  {author} {\bibinfo {author} {\bibfnamefont {S.}~\bibnamefont
			{Washburn}}, \bibinfo {author} {\bibfnamefont {R.~A.}\ \bibnamefont {Webb}},
		\bibinfo {author} {\bibfnamefont {E.}~\bibnamefont {Mendez}}, \bibinfo
		{author} {\bibfnamefont {L.}~\bibnamefont {Chang}}, \ and\ \bibinfo {author}
		{\bibfnamefont {L.}~\bibnamefont {Esaki}},\ }\href@noop {} {\bibfield
		{journal} {\bibinfo  {journal} {Physical Review B}\ }\textbf {\bibinfo
			{volume} {33}},\ \bibinfo {pages} {8848} (\bibinfo {year}
		{1986})}\BibitemShut {NoStop}%
	\bibitem [{\citenamefont {Kvon}\ \emph {et~al.}(2008)\citenamefont {Kvon},
		\citenamefont {Olshanetsky}, \citenamefont {Kozlov}, \citenamefont
		{Mikhailov},\ and\ \citenamefont {Dvoretskii}}]{Kvon2008}%
	\BibitemOpen
	\bibfield  {author} {\bibinfo {author} {\bibfnamefont {Z.~D.}\ \bibnamefont
			{Kvon}}, \bibinfo {author} {\bibfnamefont {E.}~\bibnamefont {Olshanetsky}},
		\bibinfo {author} {\bibfnamefont {D.~A.}\ \bibnamefont {Kozlov}}, \bibinfo
		{author} {\bibfnamefont {N.~N.}\ \bibnamefont {Mikhailov}}, \ and\ \bibinfo
		{author} {\bibfnamefont {S.~A.}\ \bibnamefont {Dvoretskii}},\ }\href@noop {}
	{\bibfield  {journal} {\bibinfo  {journal} {JETP Letters}\ }\textbf {\bibinfo
			{volume} {87}},\ \bibinfo {pages} {502} (\bibinfo {year} {2008})}\BibitemShut
	{NoStop}%
	\bibitem [{\citenamefont {Kvon}\ \emph {et~al.}(2011)\citenamefont {Kvon},
		\citenamefont {Olshanetsky}, \citenamefont {Novik}, \citenamefont {Kozlov},
		\citenamefont {Mikhailov}, \citenamefont {Parm},\ and\ \citenamefont
		{Dvoretsky}}]{Kvon2011}%
	\BibitemOpen
	\bibfield  {author} {\bibinfo {author} {\bibfnamefont {Z.}~\bibnamefont
			{Kvon}}, \bibinfo {author} {\bibfnamefont {E.}~\bibnamefont {Olshanetsky}},
		\bibinfo {author} {\bibfnamefont {E.}~\bibnamefont {Novik}}, \bibinfo
		{author} {\bibfnamefont {D.}~\bibnamefont {Kozlov}}, \bibinfo {author}
		{\bibfnamefont {N.}~\bibnamefont {Mikhailov}}, \bibinfo {author}
		{\bibfnamefont {I.}~\bibnamefont {Parm}}, \ and\ \bibinfo {author}
		{\bibfnamefont {S.}~\bibnamefont {Dvoretsky}},\ }\href@noop {} {\bibfield
		{journal} {\bibinfo  {journal} {Physical Review B}\ }\textbf {\bibinfo
			{volume} {83}},\ \bibinfo {pages} {193304} (\bibinfo {year}
		{2011})}\BibitemShut {NoStop}%
	\bibitem [{\citenamefont {Olshanetsky}\ \emph {et~al.}(2012)\citenamefont
		{Olshanetsky}, \citenamefont {Kvon}, \citenamefont {Mikhailov}, \citenamefont
		{Novik}, \citenamefont {Parm},\ and\ \citenamefont
		{Dvoretsky}}]{Olshanetsky2012}%
	\BibitemOpen
	\bibfield  {author} {\bibinfo {author} {\bibfnamefont {E.}~\bibnamefont
			{Olshanetsky}}, \bibinfo {author} {\bibfnamefont {Z.}~\bibnamefont {Kvon}},
		\bibinfo {author} {\bibfnamefont {N.}~\bibnamefont {Mikhailov}}, \bibinfo
		{author} {\bibfnamefont {E.}~\bibnamefont {Novik}}, \bibinfo {author}
		{\bibfnamefont {I.}~\bibnamefont {Parm}}, \ and\ \bibinfo {author}
		{\bibfnamefont {S.}~\bibnamefont {Dvoretsky}},\ }\href@noop {} {\bibfield
		{journal} {\bibinfo  {journal} {Solid State Communications}\ }\textbf
		{\bibinfo {volume} {152}},\ \bibinfo {pages} {265} (\bibinfo {year}
		{2012})}\BibitemShut {NoStop}%
	\bibitem [{\citenamefont {Minkov}\ \emph {et~al.}(2013)\citenamefont {Minkov},
		\citenamefont {Germanenko}, \citenamefont {Rut}, \citenamefont
		{Sherstobitov}, \citenamefont {Dvoretski},\ and\ \citenamefont
		{Mikhailov}}]{Minkov2013}%
	\BibitemOpen
	\bibfield  {author} {\bibinfo {author} {\bibfnamefont {G.~M.}\ \bibnamefont
			{Minkov}}, \bibinfo {author} {\bibfnamefont {A.~V.}\ \bibnamefont
			{Germanenko}}, \bibinfo {author} {\bibfnamefont {O.}~\bibnamefont {Rut}},
		\bibinfo {author} {\bibfnamefont {A.}~\bibnamefont {Sherstobitov}}, \bibinfo
		{author} {\bibfnamefont {S.~A.}\ \bibnamefont {Dvoretski}}, \ and\ \bibinfo
		{author} {\bibfnamefont {N.~N.}\ \bibnamefont {Mikhailov}},\ }\href@noop {}
	{\bibfield  {journal} {\bibinfo  {journal} {Physical Review B}\ }\textbf
		{\bibinfo {volume} {88}},\ \bibinfo {pages} {155306} (\bibinfo {year}
		{2013})}\BibitemShut {NoStop}%
	\bibitem [{\citenamefont {Halperin}\ and\ \citenamefont
		{Rice}(1968)}]{halperin1968excitonic}%
	\BibitemOpen
	\bibfield  {author} {\bibinfo {author} {\bibfnamefont {B.}~\bibnamefont
			{Halperin}}\ and\ \bibinfo {author} {\bibfnamefont {T.}~\bibnamefont
			{Rice}},\ }\href@noop {} {\bibfield  {journal} {\bibinfo  {journal} {Solid
				State Physics}\ }\textbf {\bibinfo {volume} {21}},\ \bibinfo {pages} {115}
		(\bibinfo {year} {1968})}\BibitemShut {NoStop}%
	\bibitem [{\citenamefont {Zittartz}(1967)}]{zittartz1967theory}%
	\BibitemOpen
	\bibfield  {author} {\bibinfo {author} {\bibfnamefont {J.}~\bibnamefont
			{Zittartz}},\ }\href@noop {} {\bibfield  {journal} {\bibinfo  {journal}
			{Physical Review}\ }\textbf {\bibinfo {volume} {164}},\ \bibinfo {pages}
		{575} (\bibinfo {year} {1967})}\BibitemShut {NoStop}%
	\bibitem [{\citenamefont {Zittartz}(1968)}]{zittartz1968ctransport}%
	\BibitemOpen
	\bibfield  {author} {\bibinfo {author} {\bibfnamefont {J.}~\bibnamefont
			{Zittartz}},\ }\href@noop {} {\bibfield  {journal} {\bibinfo  {journal}
			{Physical Review}\ }\textbf {\bibinfo {volume} {165}},\ \bibinfo {pages}
		{605} (\bibinfo {year} {1968})}\BibitemShut {NoStop}%
	\bibitem [{\citenamefont {Neuenschwander}\ and\ \citenamefont
		{Wachter}(1990)}]{Neuenschwander1990}%
	\BibitemOpen
	\bibfield  {author} {\bibinfo {author} {\bibfnamefont {J.}~\bibnamefont
			{Neuenschwander}}\ and\ \bibinfo {author} {\bibfnamefont {P.}~\bibnamefont
			{Wachter}},\ }\href@noop {} {\bibfield  {journal} {\bibinfo  {journal} {PRB}\
		}\textbf {\bibinfo {volume} {41}},\ \bibinfo {pages} {12693} (\bibinfo {year}
		{1990})}\BibitemShut {NoStop}%
	\bibitem [{\citenamefont {Brandt}\ and\ \citenamefont
		{Chudinov}(1972)}]{brandt1972investigation}%
	\BibitemOpen
	\bibfield  {author} {\bibinfo {author} {\bibfnamefont {N.}~\bibnamefont
			{Brandt}}\ and\ \bibinfo {author} {\bibfnamefont {S.}~\bibnamefont
			{Chudinov}},\ }\href@noop {} {\bibfield  {journal} {\bibinfo  {journal}
			{Journal of Low Temperature Physics}\ }\textbf {\bibinfo {volume} {8}},\
		\bibinfo {pages} {339} (\bibinfo {year} {1972})}\BibitemShut {NoStop}%
	\bibitem [{\citenamefont {Du}\ \emph {et~al.}(2015)\citenamefont {Du},
		\citenamefont {Lou}, \citenamefont {Chang}, \citenamefont {Sullivan},\ and\
		\citenamefont {Du}}]{du2015gate}%
	\BibitemOpen
	\bibfield  {author} {\bibinfo {author} {\bibfnamefont {L.}~\bibnamefont
			{Du}}, \bibinfo {author} {\bibfnamefont {W.}~\bibnamefont {Lou}}, \bibinfo
		{author} {\bibfnamefont {K.}~\bibnamefont {Chang}}, \bibinfo {author}
		{\bibfnamefont {G.}~\bibnamefont {Sullivan}}, \ and\ \bibinfo {author}
		{\bibfnamefont {R.-R.}\ \bibnamefont {Du}},\ }\href@noop {} {\bibfield
		{journal} {\bibinfo  {journal} {arXiv preprint arXiv:1508.04509}\ } (\bibinfo
		{year} {2015})}\BibitemShut {NoStop}%
	\bibitem [{\citenamefont {Olshanetsky}\ \emph {et~al.}(2014)\citenamefont
		{Olshanetsky}, \citenamefont {Kvon}, \citenamefont {Gerasimenko},
		\citenamefont {Prudkoglyad}, \citenamefont {Pudalov}, \citenamefont
		{Mikhailov},\ and\ \citenamefont {Dvoretsky}}]{olshanetsky2014metal}%
	\BibitemOpen
	\bibfield  {author} {\bibinfo {author} {\bibfnamefont {E.}~\bibnamefont
			{Olshanetsky}}, \bibinfo {author} {\bibfnamefont {Z.~D.}\ \bibnamefont
			{Kvon}}, \bibinfo {author} {\bibfnamefont {Y.~A.}\ \bibnamefont
			{Gerasimenko}}, \bibinfo {author} {\bibfnamefont {V.}~\bibnamefont
			{Prudkoglyad}}, \bibinfo {author} {\bibfnamefont {V.~M.}\ \bibnamefont
			{Pudalov}}, \bibinfo {author} {\bibfnamefont {N.~N.}\ \bibnamefont
			{Mikhailov}}, \ and\ \bibinfo {author} {\bibfnamefont {S.}~\bibnamefont
			{Dvoretsky}},\ }\href@noop {} {\bibfield  {journal} {\bibinfo  {journal}
			{JETP Letters}\ }\textbf {\bibinfo {volume} {98}},\ \bibinfo {pages} {843}
		(\bibinfo {year} {2014})}\BibitemShut {NoStop}%
	\bibitem [{\citenamefont {Knap}\ \emph {et~al.}(2014)\citenamefont {Knap},
		\citenamefont {Sau}, \citenamefont {Halperin},\ and\ \citenamefont
		{Demler}}]{Knap2014}%
	\BibitemOpen
	\bibfield  {author} {\bibinfo {author} {\bibfnamefont {M.}~\bibnamefont
			{Knap}}, \bibinfo {author} {\bibfnamefont {J.~D.}\ \bibnamefont {Sau}},
		\bibinfo {author} {\bibfnamefont {B.~I.}\ \bibnamefont {Halperin}}, \ and\
		\bibinfo {author} {\bibfnamefont {E.}~\bibnamefont {Demler}},\ }\href@noop {}
	{\bibfield  {journal} {\bibinfo  {journal} {Physical review letters}\
		}\textbf {\bibinfo {volume} {113}},\ \bibinfo {pages} {186801} (\bibinfo
		{year} {2014})}\BibitemShut {NoStop}%
	\bibitem [{\citenamefont {Dvoretsky}\ \emph {et~al.}(2007)\citenamefont
		{Dvoretsky}, \citenamefont {Ikusov}, \citenamefont {Kvon}, \citenamefont
		{Mikhailov}, \citenamefont {Dai}, \citenamefont {Smirnov}, \citenamefont
		{Sidorov},\ and\ \citenamefont {Shvets}}]{dvoretsky2007growing}%
	\BibitemOpen
	\bibfield  {author} {\bibinfo {author} {\bibfnamefont {S.}~\bibnamefont
			{Dvoretsky}}, \bibinfo {author} {\bibfnamefont {D.}~\bibnamefont {Ikusov}},
		\bibinfo {author} {\bibfnamefont {D.~K.}\ \bibnamefont {Kvon}}, \bibinfo
		{author} {\bibfnamefont {N.}~\bibnamefont {Mikhailov}}, \bibinfo {author}
		{\bibfnamefont {N.}~\bibnamefont {Dai}}, \bibinfo {author} {\bibfnamefont
			{R.}~\bibnamefont {Smirnov}}, \bibinfo {author} {\bibfnamefont {Y.~G.}\
			\bibnamefont {Sidorov}}, \ and\ \bibinfo {author} {\bibfnamefont
			{V.}~\bibnamefont {Shvets}},\ }\href@noop {} {\bibfield  {journal} {\bibinfo
			{journal} {Optoelectronics, Instrumentation and Data Processing}\ }\textbf
		{\bibinfo {volume} {43}},\ \bibinfo {pages} {375} (\bibinfo {year}
		{2007})}\BibitemShut {NoStop}%
	\bibitem [{\citenamefont {Olshanetsky}\ \emph {et~al.}(2009)\citenamefont
		{Olshanetsky}, \citenamefont {Kvon}, \citenamefont {Entin}, \citenamefont
		{Magarill}, \citenamefont {Mikhailov}, \citenamefont {Parm},\ and\
		\citenamefont {Dvoretsky}}]{Olshanetsky2009}%
	\BibitemOpen
	\bibfield  {author} {\bibinfo {author} {\bibfnamefont {E.}~\bibnamefont
			{Olshanetsky}}, \bibinfo {author} {\bibfnamefont {Z.}~\bibnamefont {Kvon}},
		\bibinfo {author} {\bibfnamefont {M.}~\bibnamefont {Entin}}, \bibinfo
		{author} {\bibfnamefont {L.}~\bibnamefont {Magarill}}, \bibinfo {author}
		{\bibfnamefont {N.}~\bibnamefont {Mikhailov}}, \bibinfo {author}
		{\bibfnamefont {I.}~\bibnamefont {Parm}}, \ and\ \bibinfo {author}
		{\bibfnamefont {S.}~\bibnamefont {Dvoretsky}},\ }\href@noop {} {\bibfield
		{journal} {\bibinfo  {journal} {JETP letters}\ }\textbf {\bibinfo {volume}
			{89}},\ \bibinfo {pages} {290} (\bibinfo {year} {2009})}\BibitemShut
	{NoStop}%
	\bibitem [{\citenamefont {Kirichenko}\ \emph {et~al.}(2005)\citenamefont
		{Kirichenko}, \citenamefont {Kornilov},\ and\ \citenamefont
		{Pudalov}}]{kirichenko2005properties}%
	\BibitemOpen
	\bibfield  {author} {\bibinfo {author} {\bibfnamefont {A.}~\bibnamefont
			{Kirichenko}}, \bibinfo {author} {\bibfnamefont {A.}~\bibnamefont
			{Kornilov}}, \ and\ \bibinfo {author} {\bibfnamefont {V.}~\bibnamefont
			{Pudalov}},\ }\href@noop {} {\bibfield  {journal} {\bibinfo  {journal}
			{Instruments and Experimental Techniques}\ }\textbf {\bibinfo {volume}
			{48}},\ \bibinfo {pages} {813} (\bibinfo {year} {2005})}\BibitemShut
	{NoStop}%
	\bibitem [{\citenamefont {Gusev}\ \emph {et~al.}(2012)\citenamefont {Gusev},
		\citenamefont {Olshanetsky}, \citenamefont {Kvon}, \citenamefont {Levin},
		\citenamefont {Mikhailov},\ and\ \citenamefont {Dvoretsky}}]{Gusev2012}%
	\BibitemOpen
	\bibfield  {author} {\bibinfo {author} {\bibfnamefont {G.}~\bibnamefont
			{Gusev}}, \bibinfo {author} {\bibfnamefont {E.}~\bibnamefont {Olshanetsky}},
		\bibinfo {author} {\bibfnamefont {Z.}~\bibnamefont {Kvon}}, \bibinfo {author}
		{\bibfnamefont {A.}~\bibnamefont {Levin}}, \bibinfo {author} {\bibfnamefont
			{N.}~\bibnamefont {Mikhailov}}, \ and\ \bibinfo {author} {\bibfnamefont
			{S.}~\bibnamefont {Dvoretsky}},\ }\href@noop {} {\bibfield  {journal}
		{\bibinfo  {journal} {Physical review letters}\ }\textbf {\bibinfo {volume}
			{108}},\ \bibinfo {pages} {226804} (\bibinfo {year} {2012})}\BibitemShut
	{NoStop}%
	\bibitem [{\citenamefont {Krishtopenko}\ \emph {et~al.}(2016)\citenamefont
		{Krishtopenko}, \citenamefont {Yahniuk}, \citenamefont {But}, \citenamefont
		{Gavrilenko}, \citenamefont {Knap},\ and\ \citenamefont
		{Teppe}}]{krishtopenko2016pressure}%
	\BibitemOpen
	\bibfield  {author} {\bibinfo {author} {\bibfnamefont {S.}~\bibnamefont
			{Krishtopenko}}, \bibinfo {author} {\bibfnamefont {I.}~\bibnamefont
			{Yahniuk}}, \bibinfo {author} {\bibfnamefont {D.}~\bibnamefont {But}},
		\bibinfo {author} {\bibfnamefont {V.}~\bibnamefont {Gavrilenko}}, \bibinfo
		{author} {\bibfnamefont {W.}~\bibnamefont {Knap}}, \ and\ \bibinfo {author}
		{\bibfnamefont {F.}~\bibnamefont {Teppe}},\ }\href@noop {} {\bibfield
		{journal} {\bibinfo  {journal} {Physical Review B}\ }\textbf {\bibinfo
			{volume} {94}},\ \bibinfo {pages} {245402} (\bibinfo {year}
		{2016})}\BibitemShut {NoStop}%
	\bibitem [{\citenamefont {Shklovskii}\ and\ \citenamefont
		{Efros}(2013)}]{shklovskii2013electronic}%
	\BibitemOpen
	\bibfield  {author} {\bibinfo {author} {\bibfnamefont {B.~I.}\ \bibnamefont
			{Shklovskii}}\ and\ \bibinfo {author} {\bibfnamefont {A.~L.}\ \bibnamefont
			{Efros}},\ }\href@noop {} {\emph {\bibinfo {title} {Electronic properties of
				doped semiconductors}}},\ Vol.~\bibinfo {volume} {45}\ (\bibinfo  {publisher}
	{Springer Science \& Business Media},\ \bibinfo {year} {2013})\BibitemShut
	{NoStop}%
	\bibitem [{\citenamefont {Gusev}\ \emph {et~al.}(2013)\citenamefont {Gusev},
		\citenamefont {Kvon}, \citenamefont {Olshanetsky}, \citenamefont {Levin},
		\citenamefont {Krupko}, \citenamefont {Portal}, \citenamefont {Mikhailov},\
		and\ \citenamefont {Dvoretsky}}]{Gusev2013a}%
	\BibitemOpen
	\bibfield  {author} {\bibinfo {author} {\bibfnamefont {G.}~\bibnamefont
			{Gusev}}, \bibinfo {author} {\bibfnamefont {Z.}~\bibnamefont {Kvon}},
		\bibinfo {author} {\bibfnamefont {E.}~\bibnamefont {Olshanetsky}}, \bibinfo
		{author} {\bibfnamefont {A.}~\bibnamefont {Levin}}, \bibinfo {author}
		{\bibfnamefont {Y.}~\bibnamefont {Krupko}}, \bibinfo {author} {\bibfnamefont
			{J.}~\bibnamefont {Portal}}, \bibinfo {author} {\bibfnamefont
			{N.}~\bibnamefont {Mikhailov}}, \ and\ \bibinfo {author} {\bibfnamefont
			{S.}~\bibnamefont {Dvoretsky}},\ }\href@noop {} {\bibfield  {journal}
		{\bibinfo  {journal} {arXiv preprint arXiv:1308.4356}\ } (\bibinfo {year}
		{2013})}\BibitemShut {NoStop}%
	\bibitem [{\citenamefont {Daly}\ \emph {et~al.}(1996)\citenamefont {Daly},
		\citenamefont {Dalton}, \citenamefont {Lakrimi}, \citenamefont {Mason},
		\citenamefont {Nicholas}, \citenamefont {Van~der Burgt}, \citenamefont
		{Walker}, \citenamefont {Maude},\ and\ \citenamefont
		{Portal}}]{daly1996zero}%
	\BibitemOpen
	\bibfield  {author} {\bibinfo {author} {\bibfnamefont {M.}~\bibnamefont
			{Daly}}, \bibinfo {author} {\bibfnamefont {K.}~\bibnamefont {Dalton}},
		\bibinfo {author} {\bibfnamefont {M.}~\bibnamefont {Lakrimi}}, \bibinfo
		{author} {\bibfnamefont {N.}~\bibnamefont {Mason}}, \bibinfo {author}
		{\bibfnamefont {R.}~\bibnamefont {Nicholas}}, \bibinfo {author}
		{\bibfnamefont {M.}~\bibnamefont {Van~der Burgt}}, \bibinfo {author}
		{\bibfnamefont {P.}~\bibnamefont {Walker}}, \bibinfo {author} {\bibfnamefont
			{D.}~\bibnamefont {Maude}}, \ and\ \bibinfo {author} {\bibfnamefont
			{J.}~\bibnamefont {Portal}},\ }\href@noop {} {\bibfield  {journal} {\bibinfo
			{journal} {Physical Review B}\ }\textbf {\bibinfo {volume} {53}},\ \bibinfo
		{pages} {R10524} (\bibinfo {year} {1996})}\BibitemShut {NoStop}%
	\bibitem [{\citenamefont {Raikh}\ and\ \citenamefont
		{Glazman}(1995)}]{raikh1995negative}%
	\BibitemOpen
	\bibfield  {author} {\bibinfo {author} {\bibfnamefont {M.}~\bibnamefont
			{Raikh}}\ and\ \bibinfo {author} {\bibfnamefont {L.}~\bibnamefont
			{Glazman}},\ }\href@noop {} {\bibfield  {journal} {\bibinfo  {journal}
			{Physical review letters}\ }\textbf {\bibinfo {volume} {75}},\ \bibinfo
		{pages} {128} (\bibinfo {year} {1995})}\BibitemShut {NoStop}%
	\bibitem{supp} See Supplemental Material at [URL will be inserted by publisher] for more details on magnetoresistance evolution with pressure.
	\bibitem [{\citenamefont {Voiskovskii}\ and\ \citenamefont
		{Pudalov}(1995)}]{voiskovskii1995negative}%
	\BibitemOpen
	\bibfield  {author} {\bibinfo {author} {\bibfnamefont {A.}~\bibnamefont
			{Voiskovskii}}\ and\ \bibinfo {author} {\bibfnamefont {V.}~\bibnamefont
			{Pudalov}},\ }\href@noop {} {\bibfield  {journal} {\bibinfo  {journal} {JETP
				Letters}\ }\textbf {\bibinfo {volume} {62}},\ \bibinfo {pages} {947}
		(\bibinfo {year} {1995})}\BibitemShut {NoStop}%
	\bibitem [{\citenamefont {Minkov}\ \emph {et~al.}(2012)\citenamefont {Minkov},
		\citenamefont {Germanenko}, \citenamefont {Rut}, \citenamefont
		{Sherstobitov}, \citenamefont {Dvoretski},\ and\ \citenamefont
		{Mikhailov}}]{Minkov2012}%
	\BibitemOpen
	\bibfield  {author} {\bibinfo {author} {\bibfnamefont {G.}~\bibnamefont
			{Minkov}}, \bibinfo {author} {\bibfnamefont {A.}~\bibnamefont {Germanenko}},
		\bibinfo {author} {\bibfnamefont {O.}~\bibnamefont {Rut}}, \bibinfo {author}
		{\bibfnamefont {A.}~\bibnamefont {Sherstobitov}}, \bibinfo {author}
		{\bibfnamefont {S.}~\bibnamefont {Dvoretski}}, \ and\ \bibinfo {author}
		{\bibfnamefont {N.}~\bibnamefont {Mikhailov}},\ }\href@noop {} {\bibfield
		{journal} {\bibinfo  {journal} {Physical Review B}\ }\textbf {\bibinfo
			{volume} {85}},\ \bibinfo {pages} {235312} (\bibinfo {year}
		{2012})}\BibitemShut {NoStop}%
\end{thebibliography}
\cleardoublepage
\appendix*
\section{Modeling  the magnetoconductivity  for positive $V_{G}$.}
In this section we present a few additional pictures, which illustrate results
of application of the two-band Drude model to the MC data for positive gate voltages.
\begin{figure}[h]
	\includegraphics[width=8.5cm]{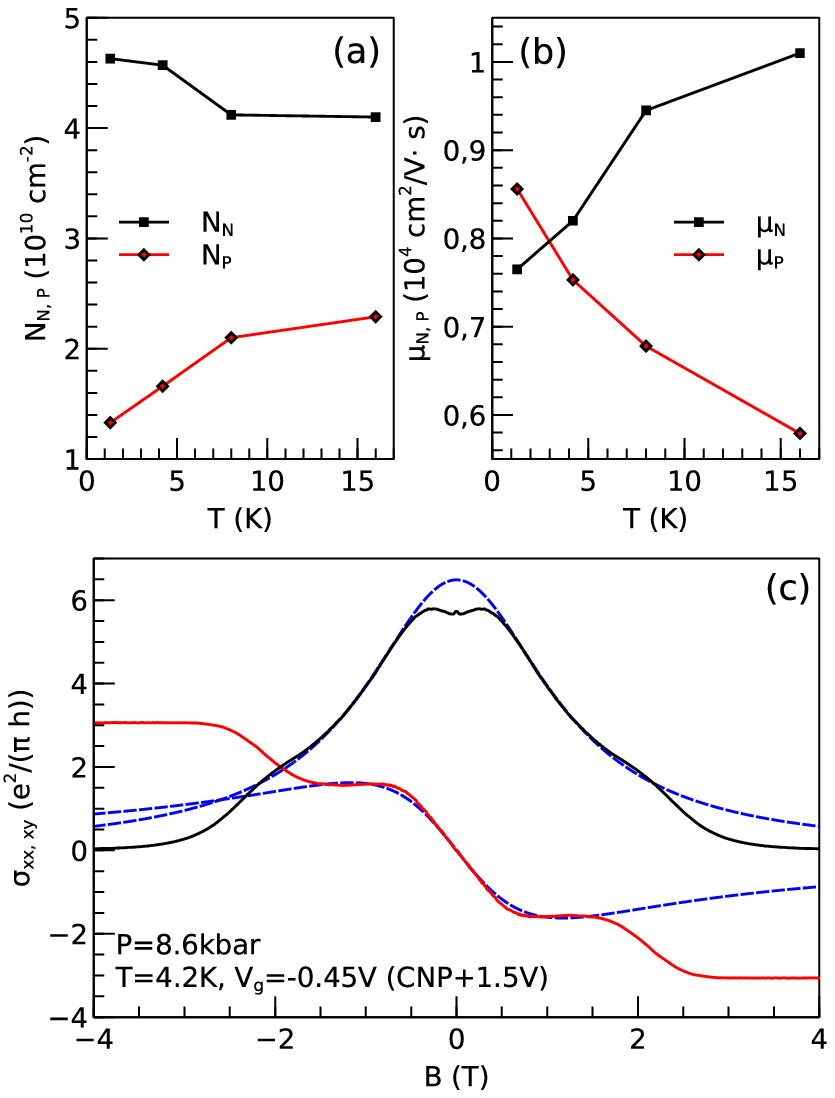}
	\caption{\label{figmun8_6}(Color online) Results of data fitting with two-band Drude model for $P=8.6$\,kbar, $V_{G}=V_{CNP}+1.5V$. (a) Electron ($N_N$) and hole ($N_P$) concentrations as a function of temperature. (b) Electron ($\mu_N$) and hole ($\mu_P$) mobilities as a function of temperature. (c) Example of experimental and theoretical curves at a particular temperature T=4.2K.}
\end{figure}
\begin{figure}[h]
	\includegraphics[width=8.5cm]{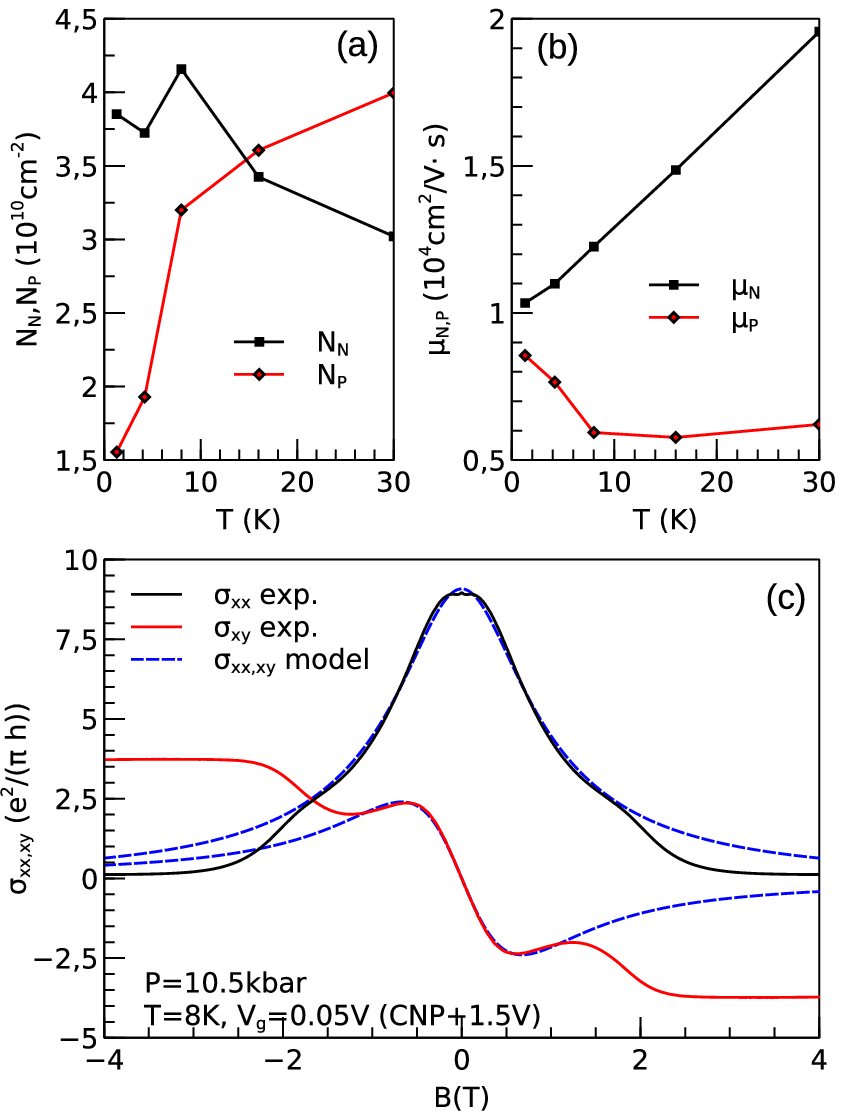}
	\caption{\label{figmun10_5}(Color online) Results of data fitting with two-band Drude model for $P=10.5$\,kbar, $V_{G}=V_{CNP}+1.5V$. (a) Electron ($N_N$) and hole ($N_P$) concentrations as a function of temperature. (b) Electron ($\mu_N$) and hole ($\mu_P$) mobilities as a function of temperature. (c) Example of experimental and theoretical curves at a particular temperature T=8K.}
\end{figure}
Fitting procedure was as follows. In order to reduce the number of fitting parameters, at the first step, we determined the zero field slope of $\sigma_{xy}(B)$ . Secondly, we determined the zero field conductivity $\sigma_{xx}(0)$ by quadratic extrapolation of $\sigma_{xx}(B)$ dependence in moderate fields (typically $0.3-0.5 T$) to zero. Direct determination of the Drude $\sigma^D_{xx}(0)$ is rather difficult due to the presence of interference-induced corrections to conductivity at low temperatures and in low fields. Nevertheless, these contributions don't affect the results in higher fields, may be reasonably excluded in weak field by parabolic extrapolation of the semiclassical fit to $B=0$; as will be shown further, both WAL  and WL  contributions may be then well fitted using the fitting parameters deduced at the first step. With these two quantities known, at the first step only two fitting parameters are left, i.e., electron and hole mobilities.

As one can see on Figs.~\ref{figmun8_6}(c) and \ref{figmun10_5}(c)  the model curves reproduce the low field behavior of both $\sigma_{xx}$ and $\sigma_{xy}$ with a reasonable set of parameters $\{N_N, N_P, \mu_N, \mu_P\}$.  The fit is successful  for all pressures and temperatures as long as we consider positive gate voltages, far enough from CNP. Though there is some scattering of parameters (Fig.~12a), taken separately, each parameter has a feasible value for our system, but the whole set and temperature evolution of parameters appear to contradict expectations  based on conventional models for transport in a 2D (semi) metallic systems.  This can be clearly seen on Figs. 11 and 12 (a,b). Indeed, (i) the temperature dependencies of electron and hole concentrations and mobilities are opposite, (ii) the ratio of electron to hole mobility is unrealistically low. The electronic effective mass at ambient pressure was found to be an order of magnitude larger than that of the holes\cite{Minkov2013}, therefore, one should expect much larger difference in the carrier mobility than the one, determined from our fitting parameters (see Fig.~12b). As was already mentioned in the main text, the small difference in mobilities  also contradicts the results of earlier measurements with semimetallic HgTe QWs at ambient pressure\cite{Olshanetsky2012, Minkov2013}.

\begin{figure}[b]
	\includegraphics[width=8.5cm]{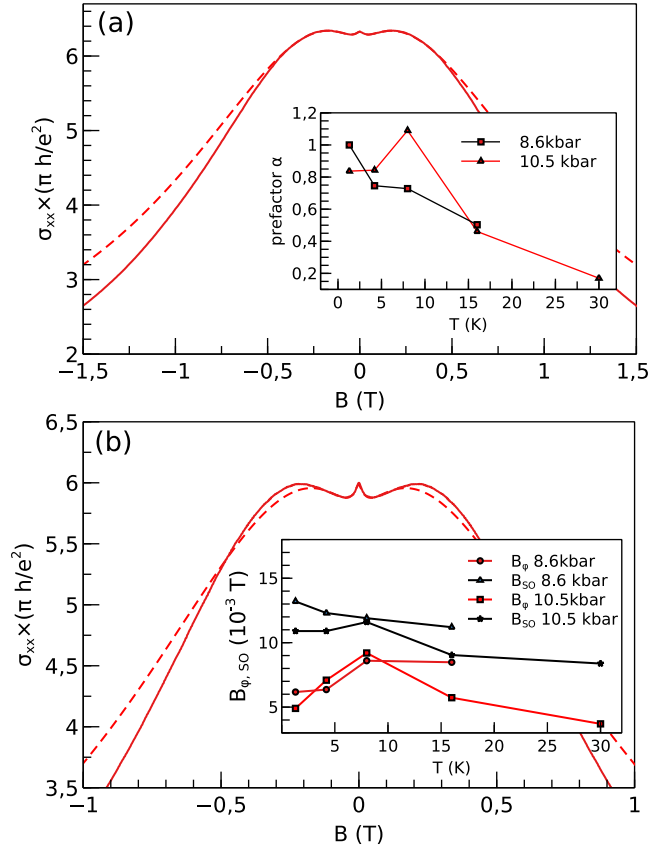}
	\caption{\label{figwalfit}(Color online) Results of data fitting with combined Drude+HLN model for (a) $P=10.5$\,kbar and (b) $8.6$\,kbar, $V_{G}=V_{CNP}+1.5V$. The inset on pannel (a) shows prefactor $\alpha$ of HLN model as a function of temperature for both 10.5\,kbar and 8.6\,kbar pressure values, $V_{G}=V_{CNP}+1.5V$. The inset on pannel (b) shows inelastic and spin-orbit decoherence fields, $B_{\varphi}$ and $B_{SO}$, vs. T.}
\end{figure}
Despite low reliability of semiclassical parameters, they still can be used to calculate the WL and WAL corrections using conventional Hikami-Larkin-Nagaoka (HLN) formula:
\begin{align*}
\Delta&\sigma(B)=-\dfrac{\alpha e^2}{2\pi^2\hbar}\left[\Psi\left(\dfrac{1}{2}+\dfrac{1}{x}\right)-\Psi\left(\dfrac{1}{2}+\dfrac{\beta_{s1}}{x}\right)\right.+\\
&\left.+\dfrac{1}{2}\Psi\left(\dfrac{1}{2}+\dfrac{\beta_{\phi}}{x}\right)-\dfrac{1}{2}\Psi\left(\dfrac{1}{2}+\dfrac{\beta_{s2}}{x}\right)\right],\,x=\dfrac{B}{B_{tr}}\\
\beta_{s1}&=\dfrac{B_{\phi}+B_{so}}{B_{tr}},\,
\beta_{s2}=\dfrac{B_{\phi}+2B_{so}}{B_{tr}},\,
\beta_{\phi}=\dfrac{B_{\phi}}{B_{tr}}
\end{align*}
A couple of examples of combined Drude and HLN fits is shown on Fig.~\ref{figwalfit}. Corresponding model parameters as a function of temperature are shown on the inset on the same figure.

As one can see,  the low field MC may be  fitted  with conventional WL model with a reasonable for our system set of parameters, but some of their temperature dependencies  look  rather scattered and, therefore,  not very reliable.
Taking into account  a questionable validity of the semiclassical background,  we  expect to obtain not a quantitatively correct temperature dependence of the WAL parameters, but just their order of magnitude.
As a result, we  associate the region of positive gate voltages with semiclassical domain and  relate the low field MC with WL-WAL corrections only  at a qualitative level. More detailed analysis requires an adequate theory.
\end{document}